\pdfoutput=1
\documentclass[twocolumn,showpacs,preprintnumbers,amsmath,amssymb,prb]{revtex4}
\usepackage{graphicx}
\usepackage{dcolumn}
\usepackage{bm}
\usepackage[tight]{subfigure}
\usepackage{amsmath}
\usepackage{verbatim}
\usepackage{color}

\graphicspath{{figs/}{}}
\begin{document}

\title{Manipulation of organic polyradicals in a single-molecule transistor}

\author{J. Fock$^{(1)}$, M. Leijnse$^{(2)}$, K. Jennum$^{(1)}$, A. S. Zyazin$^{(3)}$, J. Paaske$^{(2)}$,\\
P. Hedeg{\aa}rd$^{(2)}$, M. Br{\o}ndsted Nielsen$^{(1)}$, H. S. J. {van der Zant}$^{(3)}$}

\affiliation{
(1) Nano-Science Center \& Department of Chemistry,
    University of Copenhagen,
    DK-2100~Copenhagen \O, Denmark\\
(2) Nano-Science Center \& Niels Bohr Institute,
    University of Copenhagen,
    2100~Copenhagen \O, Denmark \\
(3) Kavli Institute of Nanoscience,
    Delft University of Technology,
    2600 GA, The Netherlands \\
}

\begin{abstract}
Inspired by cotunneling spectroscopy of spin-states in a single OPE5-based molecule, we investigate the prospects for electric control of magnetism in purely organic molecules contacted in a three-terminal geometry. Using the gate electrode, the molecule is reversibly switched between three different redox states, with magnetic spectra revealing both ferromagnetic and antiferromagnetic exchange couplings on the molecule. These observations are shown to be captured by an effective low-energy Heisenberg model, which we substantiate microscopically by a simple valence bond description of the molecule. These preliminary findings suggest an interesting route towards functionalized all-organic molecular magnetism.
\end{abstract}


\maketitle
\section{Introduction}
Single molecules are of interest as active electronic components because of the small size, the possibility of self-assembly, and because of their chemical versatility. As a consequence, the study of single-molecule transport~\cite{Joachim00, Nitzan03} is attracting massive interest. Gated three-terminal setups (single-molecule transistors)~\cite{Park00, Park02, Kubatkin03} are particularly useful, since such a geometry allows detailed spectroscopy and control of the molecular device~\cite{Osorio07b}. Especially interesting is the prospect of exploiting the spin degree of freedom in magnetic molecules, either to interact with and control the charge- or spin currents through the system (single-molecule spintronics)~\cite{Bogani08, Ruben08, Osorio10, Zyazin10, Sanvito11}, or to store and manipulate quantum information (single-molecule spin qubits)~\cite{Lehmann07b}.
In both cases, electric control of the spin states is highly desirable since, contrary to magnetic fields, these can be applied locally and varied relatively fast. Arrays of magnetic atoms, studied with spin-polarized STM, were found to be antiferromagnetically coupled~\cite{Loth12}, but ferromagnetic coupling is often more desirable since a larger net magnetic moment allows easier manipulation and readout~\cite{Chappert07}.

\begin{figure}[t!]
  \includegraphics[width=1\linewidth]{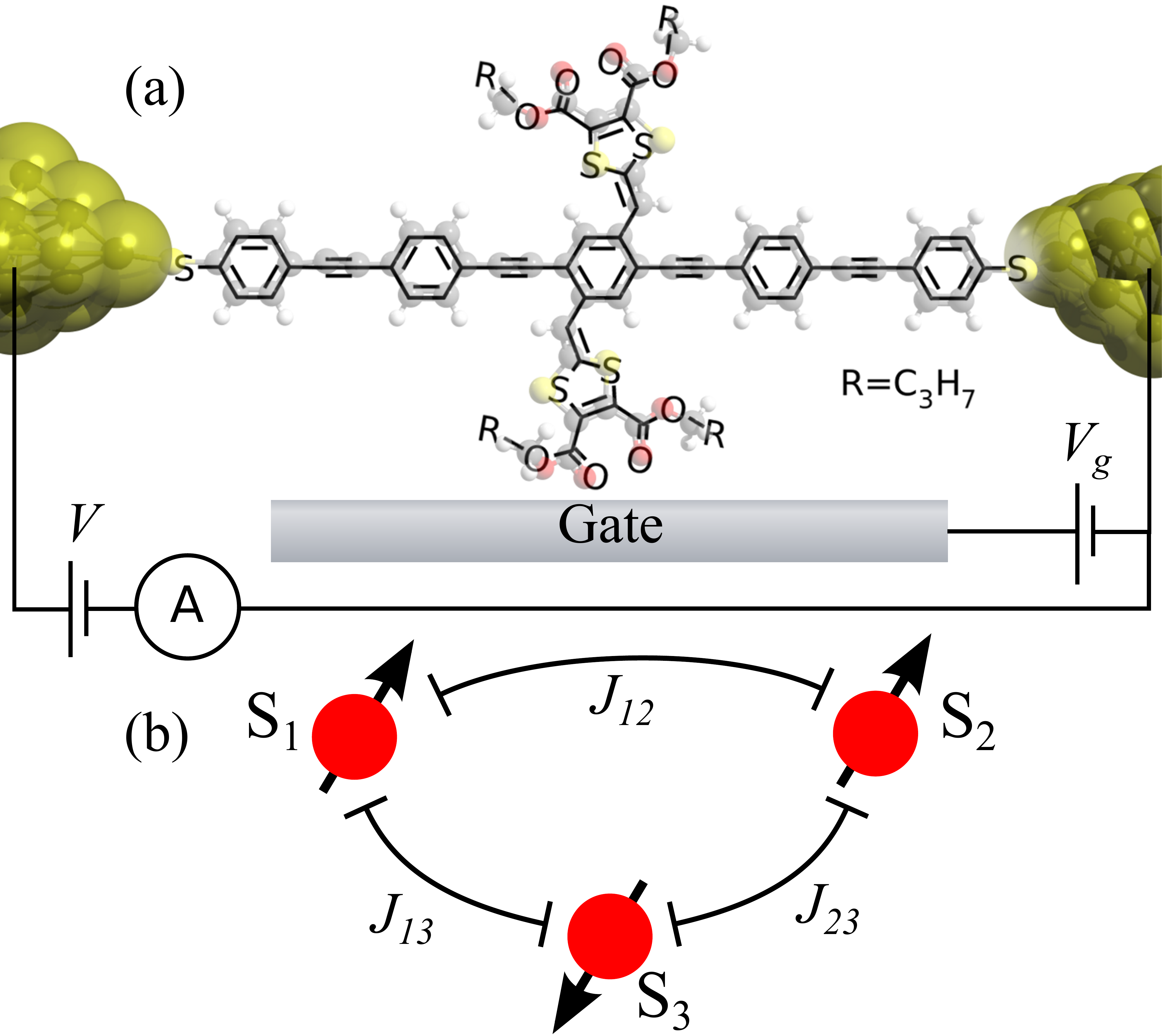}	
	\caption{\label{fig:1}
	(Color online) (a) Structure of the cruciform molecule and a sketch of the measurement setup.
	(b) Molecular spin model of three coupled spins with spin $1/2$.
	$\mathbf{S}_1$ and $\mathbf{S}_2$ are ferromagnetically
	coupled, $J_{12} < 0$, and both are antiferromagnetically coupled to $\mathbf{S}_3$ with
	$J_{13} \approx J_{23} > 0$. Increasing the gate voltage removes first $\mathbf{S}_3$
	and then $\mathbf{S}_2$.
	}
\end{figure}

Here, we discuss the prospect for electrical manipulation of spin states within a purely organic molecule, where the radicals of the charged molecule can be coupled antiferromagnetically or ferromagnetically depending on the relative location of the added charges. This is inspired by low-temperature transport measurements on a single Coulomb blockaded device, based on a molecular "cruciform": thiol end-capped penta phenylene ethynylene (OPE5) with a vertically disposed redox-active tetrathiafulvalene (TTF) unit~\cite{Jennum10}, see Fig.~\ref{fig:1}(a).
As we show below, the three-terminal transport data reveal two ferromagnetically coupled spins, $\mathbf{S}_1$ and $\mathbf{S}_2$, see Fig.~\ref{fig:1}(b). Adjusting the gate voltage to remove an electron from the molecule introduces an additional spin, $\mathbf{S}_3$, which modifies the effective interaction of $\mathbf{S}_1$ and $\mathbf{S}_2$ by coupling to them antiferromagnetically. This system is analogous to the theoretical suggestion in Refs.~\onlinecite{Lehmann07b, Lehmann09}, where $\mathbf{S}_1$ and $\mathbf{S}_2$ were used as spin qubits and two-qubit operations could be performed using the gate to control their interaction by adding/removing a third spin, $\mathbf{S}_3$.
Whereas the original proposal of Refs.~\onlinecite{Lehmann07b, Lehmann09} involved an intrinsically magnetic polyoxometalate molecule, we argue here that this can be achieved also in an all-organic molecule where a ferromagnetic ground state can be induced by gate-controlled oxidation.
Thus, the gated junction allows reversible switching of the molecule between different polyradical states.

\section{Methods}
The device is made by electromigration~\cite{Park99elmig} of a thin gold wire in a solution of the molecules, using a feedback mechanism combined with self-breaking to avoid spurious gold particle formation in the junction~\cite{Strachan05, ONeill05, Osorio08rev}. The current, $I$, was measured as a function of source drain voltage, $V$, and gate voltage, $V_g$, at a temperature $T = 2$~K. The differential conductance, $dI/dV$, was obtained from pointwise linear fits of the $I(V)$ data.

\begin{figure*}[t!]
  \includegraphics[width=\linewidth]{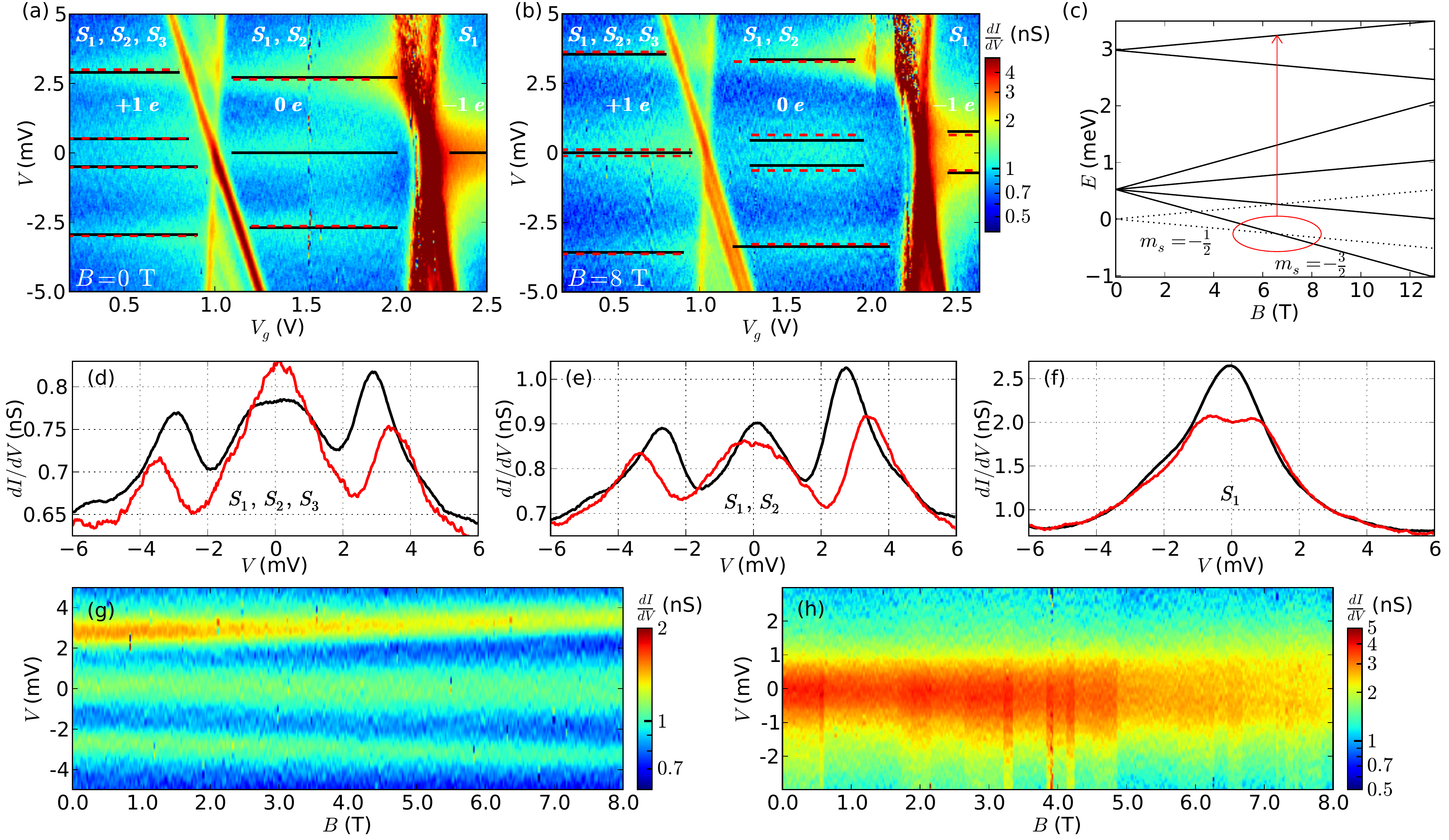}	
	\caption{\label{fig:2}
	(Color online)
	(a) $dI/dV$ measured as a function of $V$ and $V_g$ at zero magnetic field.
	(b) Same as (a), but in a magnetic field $B = 8$\,T.
	    The red dashed lines represent the excitation energies extracted from the cotunneling spectrum and
	    black solid lines are the best fit to the model Hamiltonian~(\ref{eq:hamiltonian}).
	(c) Calculated excitation spectrum [eigenenergies of Eq.~(\ref{eq:hamiltonian})]
	    as a function of magnetic field in the leftmost diamond.
	(d) $dI/dV$ as a function of $V$, obtained by averaging $dI/dV (V, V_g)$ between $V_g = 0.2$\,V and $V_g = 0.4$\,V (leftmost diamond).
	    Black (red) lines are taken at $B=0$ ($B=8$\,).
	(e) and (f) Same as (d), but obtained by averaging between $V_g = 1.4$\,V and $V_g = 1.5$\,V (leftmost diamond)
 	    and between $V_g = 2.4$\,V and $V_g = 2.5$\,V (rightmost diamond), respectively.
        (g) and (h) is magnetic field scan at $V_g=1.6$\,V (triplet ground state) and 2.6\,V  (doublet ground state) respectively.
	}
\end{figure*}

Due to intentionally low junction fabrication yields to avoid formation of junctions comprising several molecules contacted in parallel, we only obtained a single device with sufficient gate coupling ($\alpha = 0.017$) and sharp enough resonances to allow detailed spectroscopy.
In the two-dimensional conductance map presented in Fig.~\ref{fig:2}(a) and (b), gate-dependent resonances indicate when $V$ is large enough to energetically allow electrons to be added to or removed from the molecule by sequential tunneling~\cite{Osorio08rev}.

Below these lines, within the so-called Coulomb diamonds, single-electron transport is suppressed and a fixed charge is stabilized by Coulomb blockade. The two crossing points at $V = 0$ represent charge degeneracies, and by increasing $V_g$ going from left to right in Fig.~\ref{fig:2}(a) and (b) a total of two electrons are added to the molecule.
Inside the Coulomb diamonds transport takes place through coherent many-electron
cotunneling processes~\cite{DeFranceschi01}, giving rise to horizontal conductance features, which can be used to extract information about the molecular spin excitations~\cite{Osorio10, Zyazin10}. In addition, the Kondo effect~\cite{Goldhaber98, Cronenwett98, Nygard00} gives rise to conductance peaks centered at $V = 0$.
\section{Results and discussion}
Figure~\ref{fig:2}(a) represents a measurement in zero magnetic field, while Fig.~\ref{fig:2}(b) was taken at $B = 8$\,T. The dashed horizontal lines indicate the excitation energies in each charge state. As seen from the line scans in Figs.~\ref{fig:2}(d)--(e), the excitations at finite bias-voltage appear as a splitting of a much broader Kondo-resonance, rather than proper inelastic cotunneling steps. Therefore, we choose to read off the excitation energies from the center of the peaks rather than from their inflection points~\cite{Weichselbaum09,Schmitt11}.
For the present analysis, we shall merely extract the thresholds/splittings as an estimate of the exchange couplings $J_{12}$ and $J_{13}$.

\subsection{Spin model}
From the clear observation of a Zeeman effect (Fig.~\ref{fig:2}(d)-(h)) we infer that the observed low-energy excitations must be of magnetic origin.
Since changing $V_g$ across a charge degeneracy point adds one electron at the time and we observe no spin blockade~\cite{Weinmann95}, the most
general effective three-spin model is the one sketched in Fig.~\ref{fig:1}(b), described by the Hamiltonian
\begin{align}\label{eq:hamiltonian}
	H 	&= 	J_{12} \mathbf{S}_1 \cdot \mathbf{S}_2 +
			J_{13} \mathbf{S}_1 \cdot \mathbf{S}_3 +
			J_{23} \mathbf{S}_2 \cdot \mathbf{S}_3 \nonumber \\
		&+	g \mu_B \mathbf{B}\cdot \sum_i \mathbf{S}_i,
\end{align}
expressed in terms of three distinct $S=1/2$ states, magnetic field, ${\bf B}$, the Bohr magneton, $\mu_B$,
and the electron $g$-factor.
As we will show below, in the rightmost diamond
$\mathbf{S}_2 = \mathbf{S}_3 = \mathbf{0}$, and in the middle one $\mathbf{S}_3 = \mathbf{0}$.

The model allows an accurate fit of the experimentally observed excitation energies and
their magnetic field dependence (compare black full lines and red dashed lines in
Fig~\ref{fig:2}(a) and (b)). Importantly, the model also correctly predicts which excitations are
suppressed due to spin-selection rules (see below).
The exchange couplings and electron $g$-factor are fitted to the excitations energies in
Figs.~\ref{fig:2}(a) and (b); the best fit is shown by the full black lines.
We find $J_{12} = -2.63\pm0.08$\,meV, $J_{13} \sim J_{23} = 0.35\pm 0.06$\,meV and $g = 1.37\pm0.16$.
We ascribe the reduction of the $g$-factor to a combination of the (Knight-)shift of the spin-resonance frequency due to a strong Kondo effect~\cite{Weichselbaum09} and the
aforementioned uncertainty in determining the exact excitation energies~\cite{Schmitt11}.
The fit is not very sensitive to the exact ratio of the relatively small couplings $J_{13}$ and $J_{23}$, only to their sum. 

In the rightmost Coulomb diamond the strong zero-bias Kondo peak and its splitting into peaks at $|V| = g \mu_B B$~\cite{Goldhaber98, Cronenwett98, Nygard00} in magnetic field, indicates a spin-1/2 ground state ($\mathbf{S}_1$). The molecular charge must therefore be odd, but the absolute charge state cannot be uniquely determined from the measurements. We argue below that the molecule here is singly reduced, but this is not relevant for the present discussion about the spin states.

Going to the middle diamond, one electron has been removed. The presence of a zero-bias Kondo peak still implies a spin-degenerate ground state, and removing an electron must therefore introduce an additional spin, $\mathbf{S}_2$, which couples ferromagnetically with $\mathbf{S}_1$ to form a triplet ground state, resulting in a spin-1 Kondo effect~\cite{Roch09, Parks10}. That the spin has not changed by more than $\pm 1/2$ is consistent with the absense of spin blockade~\cite{Weinmann95}. Note that, instead of the usual enhancement of the Kondo peak upon approaching the charge degeneracy point a slight depression is observed near $V_g=2.0\,V$ in Fig.~\ref{fig:2} (a).  Similar observations have been made in Ref.~\onlinecite{Yu2005}, but an explanation of this requires a more detailed understanding of the charge fluctuations in this specific molecule. In a magnetic field, the spin-1 Kondo peak splits into low lying peaks at $|V| = g \mu_B B$. Excitations to the $S_z = -1$ component of the triplet are 
suppressed by spin-selection rules and are therefore not present in the experiment. The strong inelastic cotunneling lines observed at higher bias-voltages indicate transitions to the excited singlet configuration and occur at $|V| = -J_{12} + g \mu_B B$.

In the leftmost diamond, yet another electron has been removed, and the
molecule has acquired one more spin, ($\mathbf{S}_3$). An antiferromagnetic
coupling to $\mathbf{S}_{12} =\mathbf{S}_1 + \mathbf{S}_2$ leads to a doublet ground state ($S = 1/2$)
and a quadruplet excited state ($S = 3/2$) as seen in the spectrum and magnetic
field dependence shown in Fig.~\ref{fig:2}(c). This explains the merging of the
peaks at $B = 8$\,T as apparent from Fig.~\ref{fig:2}(d); the lowest component of
the quadruplet is almost degenerate with the lowest component
of the doublet, and a new effective doublet comprising these two
states becomes the groundstate. As a result a stronger zero-bias Kondo
effect is obtained (compare red and black lines at zero bias in Fig.~\ref{fig:2}(d)),
meaning that this finite-field doublet is better coupled to the leads than
the zero-field doublet.
A more detailed analysis shows that this can
indeed be possible, if electrons tunnel via both $\mathbf{S}_1$ and $\mathbf{S}_3$.
Similar behavior where a Kondo peak is split by interaction with another spin,
but can be restored with a magnetic field, has been observed in
STM studies of coupled magnetic and
non-magnetic atoms on surfaces~\cite{Otte09}. 
Unfortunately the device broke down, and as a consequence we do not have the full B-field dependence in this charge state, but only the conductance map at 0 and 8\,T. The full B-field dependence was however acquired at the two other charge states (Fig.~\ref{fig:2}(g-h)).

The zero-bias Kondo effect arising from the ground state doublet at $B=0$ (Fig.~\ref{fig:2}(d), black line) is markedly weaker than that observed for the single-spin in the far right
diamond (Fig.~\ref{fig:2}(f)). This can be understood by inspection of the effective three-spin ground state doublet, expressed in terms of individual spin states $|S_{1}^{z}S_{2}^{z}S_{3}^{z}\rangle$ as:
\begin{align}
|\Uparrow\rangle=&\,\,\frac{1}{\sqrt{6}}\left(
2|\uparrow\uparrow\downarrow\rangle-
|\uparrow\downarrow\uparrow\rangle-
|\downarrow\uparrow\uparrow\rangle\right)\\
|\Downarrow\rangle=&\,\,\frac{1}{\sqrt{6}}\left(
 |\uparrow\downarrow\downarrow\rangle
 +|\downarrow\uparrow\downarrow\rangle -2|\downarrow\downarrow\uparrow\rangle \right),
\end{align}
assuming for simplicity that the two antiferromagnetic couplings are equal, i.e. $J_{23} = J_{13}$. The amplitudes for flipping of this doublet by a simple exchange-cotunnelling process at either $\mathbf{S}_1$ or $\mathbf{S}_3$ are therefore reduced by Clebsch-Gordan coefficients to respectively
\begin{align}
\langle\Downarrow|S^{-}_1|\Uparrow\rangle=\frac{2}{3},\hspace{4mm}{\rm and}\hspace{4mm}
\langle\Downarrow|S^{-}_3|\Uparrow\rangle=-\frac{1}{3}.
\end{align}
The effective Kondo temperature depends exponentially on these amplitudes and is therefore significantly smaller than for a single spin. Since the measurement was carried out at $T<T_K$, a reduced Kondo temperature can easily result in a smaller zero-bias conductance. Notice that the triplet correlations between $\mathbf{S}_1$ and $\mathbf{S}_2$ makes the effective exchange-cotunneling via $\mathbf{S}_3$ effectively ferromagnetic, thus prohibiting Kondo effect altogether. This suggests that the strongest tunnel coupling to the molecule takes place via $\mathbf{S}_1$ or $\mathbf{S}_2$.

\subsection{Microscopic model}
Having rationalized the salient features of the data in terms of the Hamiltonian in eq. \ref{eq:hamiltonian}, we now proceed to give a likely microscopic understanding of this simple 3-spin model within a valence bond (VB) picture of the molecule~\cite{Shaikbook}.
This gives a description of the low-energy excitations of the molecule. The diagram in Fig.~\ref{fig:1}(a) shows one of the possible VB's in the ground state of the neutral molecule.
This configuration, however is {\em not} the ground state of the molecule between electrodes. We shall use it rather as a neutral, spin singlet, reference state. If we first add an electron we arrive at the state of the molecule in the rightmost diamond of the stability diagram. Adding the electron the following will happen:
a bond is broken, leaving a (negatively) charged carbon atom with no spin, and a neutral carbon with an unpaired spin. These two entities (charge and spin) can now move through the molecule. The charge will be particularly stable at the electrode ends of the molecule due to the attraction by image charges~\cite{Kubatkin03,Kaasbjerg2008}. Spin and charge
will not become totally independent, since both disrupt the local VB configurations, and staying close minimizes the disruption.

Going to the middle diamond, an electron will be removed from one of the two TTF-branches of the molecule, and the associated unpaired spin will predominantly move about in the same branch. Note that the overall molecule is neutral, but we propose that the lowest energy configuration has a negative charge close to one of the electrodes and a positive charge on one of the sulphur atoms in the TTFs. Once in a while the two spins can perform an exchange process, and this will result in either an antiferromagnetic or a ferromagnetic coupling, depending on which two branches of the molecule the two spins belong to.


The general rule, which determines the character of the magnetic coupling is related to the actual location of the involved spins. The carbon atoms in the molecule can be divided into two ``sublattices'', with nearest neighbor atoms belonging to different sublattices. It is a simple fact of the VB picture, that to a very good approximation, a spin belonging to a given sublattice will stay on that sublattice. The rule now is that two spins living in the same sublattice will be ferromagnetically coupled, while two spins living on different sublattices are antiferromagnetically coupled.

For the molecule we study, the ground state of the neutral molecule (the central diamond in the stability diagram) is a spin triplet. We thus require that the two spins live on the same sublattice, which leaves only two possibilities in the choice of branches with spin and charge.

\begin{figure}[t!]
  \includegraphics[width=1\linewidth,clip=True]{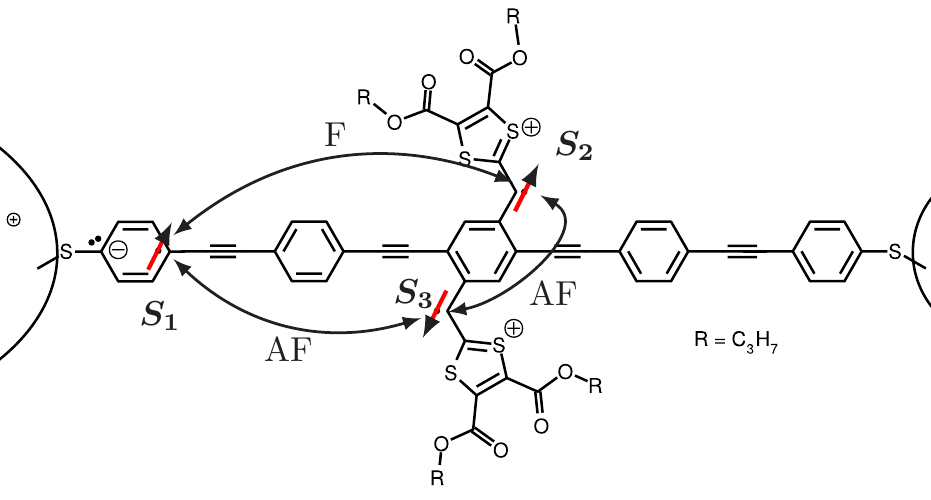}		
	\caption{\label{fig:3}
	(Color online) Tri-radical form of the molecular cruciform, corresponding to the condition in the leftmost diamond in Fig.~\ref{fig:2}.
	One charge (most likely negative), and the corresponding spin ($\mathbf{S}_1$), is localized close to one end of the OPE5 backbone.
	The vertically disposed TTF unit is oxidized twice. The corresponding spins, $\mathbf{S}_2$ ($\mathbf{S}_3$) are in a meta (ortho)-configuration
	relative to $\mathbf{S}_1$, resulting in a ferromagnetic (antiferromagnetic) coupling.
	$\mathbf{S}_2$ and $\mathbf{S}_3$ are para-coupled, resulting in an antiferromagnetic coupling.
	}
\end{figure}

Turning to the discussion of the third diamond, we extract one more electron. The charge and the spin will inhabit a third branch of the molecule. If the first two spins were ferromagnetically coupled, the third will necessarily be antiferromagnetically coupled to the first two spins, in accordance with the simplified model used in the data analysis.
Fig.~\ref{fig:3} shows the VB of this state.

What has been described in words in the previous two paragraphs can be put into rigorous mathematics in a model, where the VB system can be mapped onto an effective Heisenberg model, with as many sites, as there are coupled $p_z$ orbitals in the molecule~\cite{hedegard11}.

\section{Conclusion}
In conclusion, cotunneling spectroscopy has been used to demonstrate the realization of a series of different
spin states upon successive reduction of a single organic molecule.
If $J_{13} = J_{23}$, the Hamiltonian~(\ref{eq:hamiltonian}) is equivalent to the spin-part of Eq.~(1)
in Ref.~\onlinecite{Lehmann07b}, and thus describes a two-spin qubit system ($\mathbf{S}_1$ and $\mathbf{S}_2$),
with an interaction which is controllable by the reversible addition / removal of $\mathbf{S}_3$.
However, as long as $J_{13} \neq 0$ and $J_{23}\neq 0$, adding $\mathbf{S}_3$, and removing it again after a certain time has elapsed, results in a non-trivial change of the two-qubit state spanned by $\mathbf{S}_1$ and $\mathbf{S}_2$. Thus, this operation acts as a two-qubit gate.

Using a valence bond description of the molecule we have provided a microscopic motivation for the effective spin model, which, however, relies on interactions with image charges in the electrodes stabilizing a different charge-configuration than in the molecule in vacuum. 
Therefore, it cannot be expected that all cruciform molecule devices will behave in the same way as the single device investigated here.
However, as the data stands, lacking the full B dependence in the left-most charge state, it demonstrates that the device allowed electrical control of quantum spins in an organic molecule, with the non-trivial spin states being induced by the gate-voltage, rather than being an intrinsic property of the neutral molecule.
It is an important venue for future research to find new molecular structures where such controlled redox reactions give rise to the desired spin-structure in a more reproducible way. This could for example be achieved by adding more intrinsic molecular redox centers, which can stabilize multiple localized charges and spins without the need for interactions  with the electrodes.

\section*{Acknowledgments}
The research leading to these results has received funding from the European Union Seventh Framework Programme
(FP7/2007-2013) under
agreement no 270369 (“ELFOS”) and from FOM.
\bibliographystyle{apsrev}
\bibliography{cite_only}
\end{document}